# A 71-76 and 81-86GHz, Scaled 16-Element Transceiver Phased Array with Shared Image Selection Weaver Architecture, 25% EIRP to PDC, and Low EVM Variation

Najme Ebrahimi[1], Kamal Sarabandi[2], James Buckwalter [3]
[1,2]University of Michigan Ann Arbor
[3]University of California- Santa Barbara
[1]najme@umich.edu, [2]saraband@umich.edu, [3]buckwalter@ucsb.edu

*Abstract*—A 16-element, compact phased array transceiver is demonstrated at E-band (71-76 and 81-86GHz). A Weaver image selection architecture reduces the LO tuning range to 3GHz while covering the 10GHz band. The bidirectional and efficient implementation of Weaver architecture in scalable array is proposed using bidirectional shared image selection IF mixer. The beam is steered across +/-30° with an average 30dBm EIRP across the upper and lower bands using narrow band LO phase shifter. A 1.5 GHz modulation bandwidth for 16 QAM and 64 QAM waveform is generated with EVM better than -19 dB/ -24 dB, respectively. The 2x2 transceiver die is implemented in SiGe BiCMOS and is assembled with 16-elements while operating at a 25% EIRP/PDC efficiency.

## I. INTRODUCTION

Future 5G communication networks demand cell-to-cell backhaul and E-band (71-76 and 81-86 GHz) offers high data rates over a 2GHz channel. While E-band has lower atmospheric attenuation compared to the 60-GHz band, a relatively large array is required to compensate the path loss and a scalable array that operates over the entire E-band requires compact transmit (TX) and receive (RX) elements as well as an approach to scale to larger numbers of elements from a 2 x 2 array die. Above 60 GHz, the unit-cell size must be close to $\lambda/2$ spacing between the antennas and include local oscillator (LO) distribution, intermediate frequency (IF), I/O interfaces, transitions between antenna and IC, and heat distribution from the IC. Prior work has shown scalable arrays but not wideband transmit and receive arrays that operate over 71-76 and 81-86 with low power consumption [1-4].

The proposed approach to a scalable millimeter-wave (mm-wave) array is based on LO distribution. The LO is phase-shifted at each element to avoid large RF phase shifters in each signal path and RF gain variation over the band. The LO phase shift is produced at low frequency (19.5 GHz) and frequency-multiplied at each element. A direct conversion architecture would demand the LO shift from 71-86 GHz and require high fractional bandwidth (FBW) of 15%. Alternatively, a sliding IF approach converts the RF to a lower frequency IF and then a second mixing stage produces a second frequency conversion. If both the first and second LO are tunable, the FBW is reduced to 10%. An earlier paper presented a Weaver image selection approach that switched between the upper and lower sideband to reduce the FBW of the LO to 4% [5-6].

In this work, a scalable array is proposed based on a Weaver image selection architecture (ISA) to covers both 71-76 and 81-86GHz while reducing the LO tuning range to 4%. Shown in Fig. 1, the proposed ISA uses the lower (71-76) and upper (81-86) mm-wave bands as images of each other. In addition, the LSB and USB are switched with only one phase inversion. While the Weaver ISA could be copied across N-elements, this would suffer from scaling of the number of required I/Q mixers and bandpass filter (BPF), 2N for each of the N elements. The proposed approach shares the 1st intermediate frequency (IF1) between elements so that a single image selection stage reduces the number of I/Q mixers to 2+N and only two I/Q filter as illustrated in Fig. 1 to significantly reduce power consumption across the array TRX modes.

## II. PROPOSED N-PATH SHARED IF WEAVER ARCHITECTURE

Fig. 2 details the bidirectional IF architecture for TX and RX modes. The PA/LNA are bidirectional circuits with an input switch that is turned on in TX mode to isolate the LNA through a $\lambda/4$ line to a high impedance at the antenna. Each of the four IF signals is produced from mixing the PA/LNA port through the RF mixer to a 1st IF. In TX-mode, a common-emitter amplifier produces current injected into the ring mixer, creating an active Gilbert mixer with high conversion gain. In RX mode, the switches are passive (no dc current) to improve linearity. The input impedance at each element is $R_{sw} + Z_{IF}(\pm\omega_{IF})e^{\pm\phi_{RF}}$, where $R_{sw}$ is the switch on resistance and $Z_{IF}$ is the impedance of either the TX or RX amplifier within IF1 bandwidth. The proposed frequency plan reduces the bandwidth at IF1 and main IF to 3GHz, relaxing the matching network bandwidth, and providing flat conversion gain. The conversion gain over bandwidth between two IF and RF mixer is also shown in Fig. 2 and shows the excellent low conversion gain variation.

Each of the four IF1 signals are combined or split between the elements using a bidirectional active combiner/splitter and common LC bandpass filter. The LC filter acts as an input matching network for the common-emitter amplifier and buffer of the active splitter (TX mode) while providing optimum load for the common-base transistor (Q2) of active combiner (RX mode). The S11 is better than -10dB over

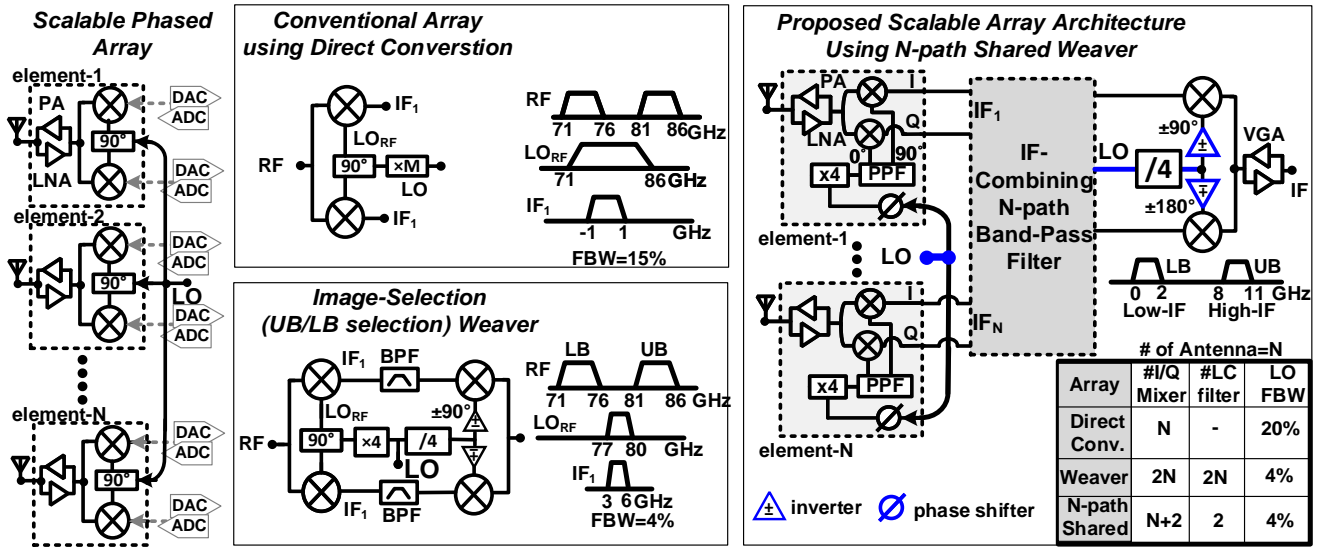

Fig. 1: Conventional direct conversion scalable phased array versus proposed scalable phased array architecture using Weaver architecture with shared image selection IF mixer and narrow-band LO phase shifter.

3GHz bandwidth at shared interface node as shown in Fig. 2. The quadrature signals are multiplied across four path at each element with the phase-shifted LO. The quadrature phase shift requires only one-stage PPF, with amplitude and phase mismatch of less than ±1dB and ±2.5 degree, respectively, over the 3GHz bandwidth, corresponding to 30-dB image rejection ratio and limiting EVM. Reduced LO FBW avoids complicated calibration circuitry to compensate the multi-stage PPF loss. More importantly, reduced LO tuning range minimizes the LO phase and amplitude error for the LO phase shifter. A 32-bit differential injection-locked oscillator (ILO) phase shifter with 1 GHz tuning range at 19.5 GHz is employed for frequency locking and phase shifting in the beam-former.

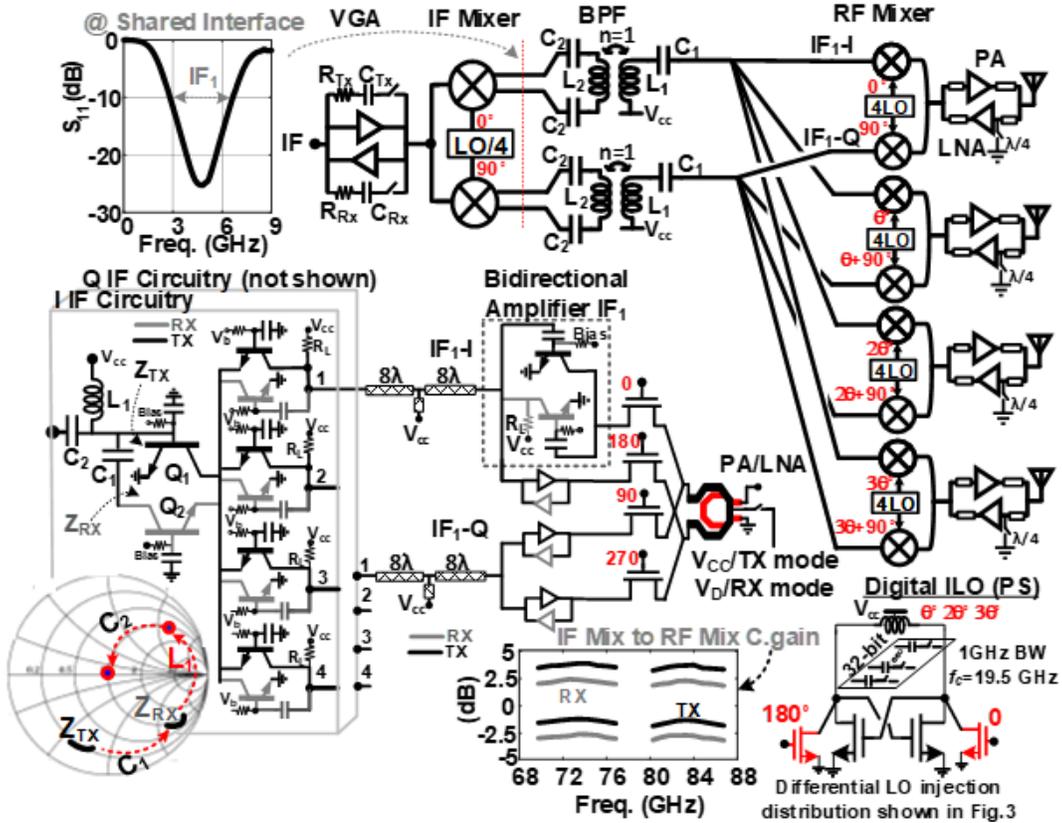

Fig. 2: The bidirectional implementation of transceiver with shared IF mixer between N-elements, the bidirectional matching for TX and RX amplifiers and mixers, the simulated conversion gain between two IF and RF mixer over two bands, and narrow-band 32-bit digital injection-locked oscillator phased shifter.

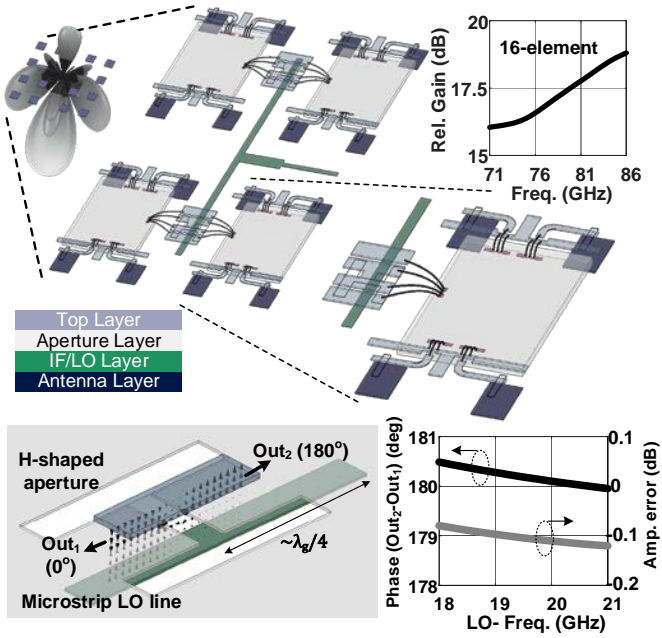
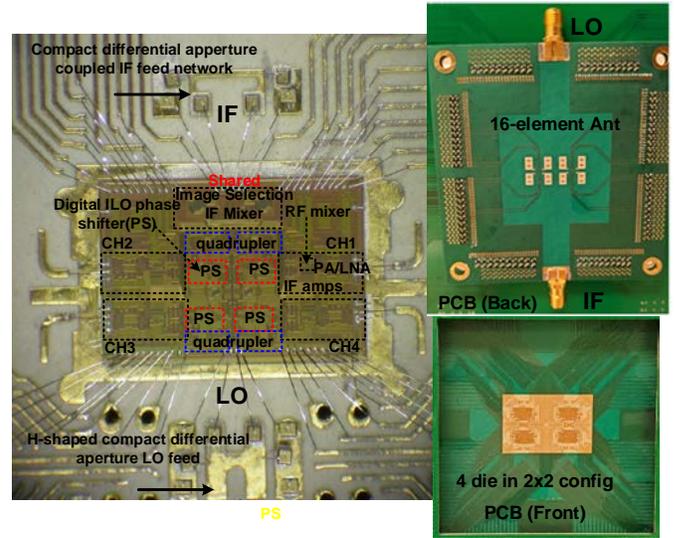

Fig. 3. The four layer 16-element antenna IC integration using aperture coupled technique for RF, and proposed compact differential aperture coupled LO feed network with simulated gain and amplitude errors.

Fig. 4. Die micrograph, and back and front-view of 4-die and 16-element antenna integration.

III. PROPOSED COMPACT SCALABLE ANTENNA-IC PACKAGING

In order to have a symmetric and compact LO distribution network between multiple dies and low-mismatch current injection between the ILO arrays, a novel compact differential feed network is proposed. On-chip balun or conventional rat-race couplers are poor candidates for compact differential feed network to meet spacing requirements between antennas. A differential, aperture-coupled balun is proposed for differential feed networks at each LO/IF port of the 2x2 array and integrated on a 4-layer PCB shown in Fig. 3 with on-PCB antennas. The electric field distribution for the proposed compact LO differential feed network is illustrated in Fig. 3. The bottom microstrip feed line is open and the aperture is approximately $\lambda/4$ from the end of the line to ensure maximum magnetic coupling. According to the electric field distribution, the signals across the aperture have the same amplitude but are 180° out of phase. An H-shaped aperture is used instead of a linear aperture. The simulated differential signal amplitude and phase mismatch are a minimum of 0.5 degree and 0.02 dB amplitude mismatch.

The 16-element antenna-IC integration uses four PCB layers. The top PCB layer routes the RF signals for the 2x2 die while eliminating the need for multi-layer vias with large loss. The aperture layers are placed between the RF and LO/IF signals, optimized for 5 mil thickness using RO 3006 substrate. To enhance the radiation gain, a 10-mil RO3003 substrate ($\varepsilon_r = 3$) is used for the antenna and aperture. The resulting aperture coupled antenna array is wideband (10-dB impedance bandwidth), exhibiting a maximum of 2dB gain variation over 71–86 GHz. The area of the 16-element array is around 11.5mm×7.1mm and fits within the $\lambda/2$ spacing limit.

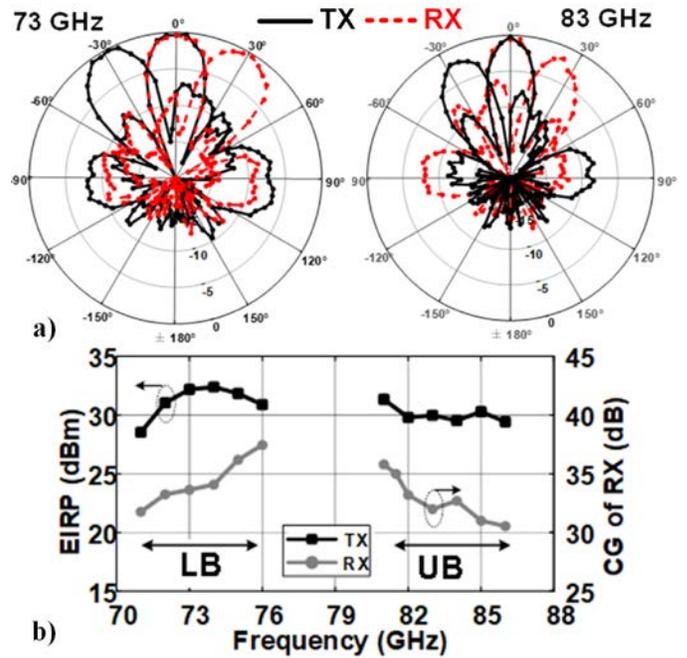

Fig. 5. a) The 16 element-measured beam-steering for two UB and LB for TX and RX mode at 73 GHz and 83 GHz, b) measured EIRP for TX mode for broadside and RX conversion gain over two UB and LB.

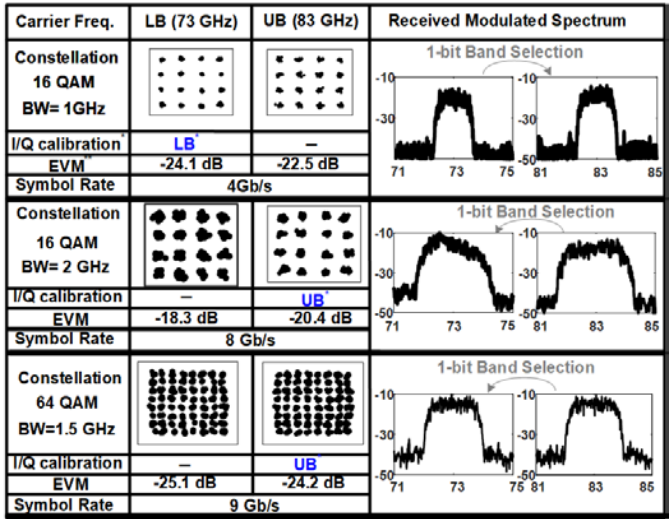

Fig. 6. The constellation diagram and EVM values for 16QAM and 64 QAM for two UB and LB. The system is optimized for calibration at one of the band, the data sent to other band with only one-bit of phase inverter switch.

## IV. MEASUREMENT RESULTS

The 16-element, E-band array is fabricated in 90nm SiGe BiCMOS. Four die are wire-bonded on a 4-layer PCB through the aperture coupled ports. The chip microphotograph is illustrated in Fig. 4 alongside the 16-element assembly. A single die is a 2x2 array occupying 4.6mm x 2.8mm. The measured antenna gain is 11 to 13 dBi due to the wideband aperture coupling technique. The array beam-steering angle for both TRX modes at both bands are shown in Fig. 5a, illustrating a maximum ±30° with 15dB peak to null ratio and 30° half-power beam-width. The $P_{sat}$ of each element is up to 8dBm and the measured TX EIRP of 28 to 32dBm over the 10GHz. The RX conversion gain is also plotted in Fig. 5b and is averaged 32dB over the band.

The EVM of the TX is measured over up to 25cm and the worst-case EVM is reported in Fig. 6 for 16 QAM and 64 QAM over both the lower and upper bands. An 8Gb/s 16 QAM modulation is transmitted with EVM less than -18 dB and for 64 QAM a 9Gb/s is achieved with EVM less than -24 dB. The EVM variation is +/- 2dB over the two bands.

The proposed wideband, scalable transceiver array is compared to prior work in Table I and demonstrates the first solution to cover both 71-76 and 81-86GHz while achieving low EVM. The power consumption of each element is 250 mW in TX mode and 160 mW in RX mode. The chip has the best EIRP/PDC over the widest RF bandwidth at 25% over 10GHz.

## V. CONCLUSION

This paper presents the first E-band scalable phased array to operate over 71-76 and 81-86GHz. It demonstrates the highest energy-efficiency at E-band with an average 25% EIRP/PDC over 10 GHz band. The results present consistent +/-30degree beam-steering for TX/RX over entire band. A novel compact differential aperture coupled LO/IF feed network for half-wavelength spacing between dies has been proposed. The system offers up to 2 GHz modulation bandwidth for 64 QAM with EVM better than -19 dB across the entire band.

## VI. ACKNOWLEDGEMENT

The authors would like to thank Nokia for the support of this research and, in particular, T. Kovarik for the support.

Table I: Performance summary and comparison of above 60 GHz scalable phased array transceiver.

| | This Work | [1] JSSC'19 | [2] ISSCC'18 | [3] ISSCC'19 | [4] ISSCC'16 |
|---|---|---|---|---|---|
| Technology | 90nm SiGe BiCMOS | 180 nm SiGe | 40 nm CMOS | 22nm FinFET | 28 nm CMOS |
| Freq. (GHz) | 71-76 81-86 | 80-100 | 57-66 | 71-76 | 57-66 |
| Element per RFIC | 4 | 16TX, 8RX | 12 | 4 | 4 |
| Array Size | 16 | 256 | 144 | 64 | 4 |
| Architecture | Sliding-IF | Direct I/Q | Direct I/Q | Direct I/Q | Direct I/Q |
| Beamforming | LO[1] | RF | RF | BB | BB |
| RF BW (GHz) | 10 | 20 | 9 | 5 | 8 |
| Phase shifter-BW (GHz) (PS/RF %) | 1[2] 10% | 20 100% | 9 100% | 2[3] 40% | 1.76[3] 22% |
| Integration | RF/LO/Analog | RF/LO/Analog BB, Memory | RF/LO/Analog BB, Memory | RF/LO/Analog | RF/LO/Analog BB |
| Antenna | On-PCB | In-package | On-PCB | On-PCB | On-PCB |
| TX $P_{sat}$ (dBm) | 8 | 8 | 5 | 8 | 9 |
| RX NF (dB) | <9 | 6.8-8 | 7 | 6 | 4.8-6.2 |
| RX conv. Gain | 32 | 80 | 23 | 37 | 62 |
| EIRP Total (dBm) | 30 | 60 | 51 | 44.4 | 24 |
| Constellation Data-rate (Gb/s) EVM (dB) | 64 QAM 9 Gb/s (-24) 16 QAM 8 Gb/s (-19) | 64 QAM[4] 18 Gb/s (-30) 16 QAM 10 Gb/s (-22) | - | 16 QAM 4 Gb/s (-25) 16 QAM 7.2 Gb/s (-20) | 64 QAM 10.5 Gb/s (-20) 16 QAM 7 Gb/s (-21) |
| EVM Variation over RF band (dB) | ± 2 dB | - | - | - | - |
| TX PDC per el.(mW) | 250 | 275 | 58 | 168 | 167 |
| RX PDC per el.(mW) | 160 | 225 | 46 | 148 | 108 |
| EIRP/PDC$_{TX}$ ×100 | 25% | 9.5% | 15% | 22% | 37%[5] |
| Die Size (mm²) | 4.6×2.8 | 8.1×4.5 | 4.68×4.48 | 2.1×2.4 | 2.6×3 |

1* Digital Injection-locked oscillator phased shifter. *2. The proposed architecture reduces LO FBW to only 3%, resulting in narrow band LO distribution (1GHz). *3: BW of I/Q baseband. *4. According to the referred paper Shahramian et.al's ISSCC 2018. *5. 4-element antenna for a single die.

## REFERENCES

[1] S. Shahramian et al., "A Fully Integrated 384-Element, 16-Tile, W-Band Phased Array With Self-Alignment and Self-Test," in *IEEE JSSC*, pp. 2419-2434, Vol. 54, No. 9, Sept. 2019.

[2] T.owlati *et al.*, "A 60GHz 144-element phased-array transceiver with 51dBm maximum EIRP and ±60° beam steering for backhaul application," *IEEE ISSCC*, pp. 66-68, 2018.

[3] S. Pellerano et al., "A scalable 71-to-76 GHz 64-element phased-array transceiver module with 2×2 direct-conversion IC in 22 nm FinFET CMOS technology," in IEEE Int. Solid-State Circuits Conf. (ISSCC) Dig. Tech. Papers, Feb. 2019, pp. 174–176.

[4] G. Mangraviti et al., "A 4-Antenna-Path Beamforming Transceiver for 60GHz Multi-Gb/s Communication in 28nm CMOS," ISSCC, pp. 246-247, Feb. 2016.

[5] N. Ebrahimi and J. F. Buckwalter, "A High-Fractional-Bandwidth, Millimeter-Wave Bidirectional Image-Selection Architecture With Narrowband LO Tuning Requirements," in *IEEE JSSC*, vol. 53, no. 8, pp. 2164-2176, Aug. 2018.

[6] N. Ebrahimi, P. Wu, M. Bagheri and J. F. Buckwalter, "A 71–86-GHz Phased Array Transceiver Using Wideband Injection-Locked Oscillator Phase Shifters," in *IEEE Transactions on Microwave Theory and Techniques*, vol. 65, no. 2, pp. 346-361, Feb. 2017.